# Evaluation of Logistic Regression Applied to Respondent-Driven Samples: Simulated and Real Data

Sandro Sperandei[1,2*], Leonardo S. Bastos[3], Marcelo Ribeiro-Alves[4], Arianne Reis[5], Francisco I. Bastos[2]

[1] Translational Health Research Institute, Western Sydney University, Australia
[2] Institute of Scientific and Technological Communication & Information in Health, Oswaldo Cruz Foundation, Brazil
[3] Scientific Computing Program, Oswaldo Cruz Foundation, Brazil
[4] National Institute of Infectious Diseases Evandro Chagas, Oswaldo Cruz Foundation, Brazil
[5] School of Health Sciences, Western Sydney University, Australia
* Corresponding author: s.martinssperandei@westernsydney.edu.au

**ABSTRACT**

**Objective:** To investigate the impact of different logistic regression estimators applied to RDS samples obtained by simulation and real data. **Methods:** Four simulated populations were created combining different connectivity models, levels of clusterization and infection processes. Each subject in the population received two attributes, only one of them related to the infection process. From each population, RDS samples with different sizes were obtained. Similarly, RDS samples were obtained from a real-world dataset. Three logistic regression estimators were applied to assess the association between the attributes and the infection status, and subsequently the observed coverage of each was measured. **Results:** The type of connectivity had more impact on estimators' performance than the clusterization level. In simulated datasets, unweighted logistic regression estimators emerged as the best option, although all estimators showed a fairly good performance. In the real dataset, the performance of weighted estimators presented some instabilities, making them a risky option. **Conclusion:** An unweighted logistic regression estimator is a reliable option to be applied to RDS samples, with similar performance to random samples and, therefore, should be the preferred option.

**Highlights:**

- Unweighted logistic regressions are the best choice for RDS studies
- RDS method can be applied to a broader spectrum of problems out of hard-to-reach populations
- Weighted estimators can be heavily affected by real-world situations

**INTRODUCTION**

Respondent-driven sampling (RDS) is a chain-referring sampling method based on the key principle that the best recruiter for a hard-to-reach, marginalized or hidden population is a member of this very population. The method's success in recruiting individuals from hard-to-reach populations is well accepted, and major international organizations have advocated its use, including the Centers for Disease Control and Prevention (Lansky & Mastro, 2008) and the World Health Organization (Johnston, Chen, Silva-Santisteban, & Raymond, 2013).

As an estimation method, it is based on the assumption that the size of an individual's contact network is related to the probability of this individual being recruited to the sample. For this reason, the accepted procedure is to weight individuals as the inverse value of their network size, resulting in individuals with smaller networks, and therefore less likely of being recruited, receiving higher weighting or importance in prevalence prediction (Gile, Johnston, & Salganik, 2015).

The performance of RDS prevalence estimators has been assessed in many studies, using different methods, particularly simulations (e.g., Goel & Salganik, 2010; Mills, Johnson, Hickman, Jones, & Colijn, 2014), with varying results. In general, studies have shown an intermediate to high performance of RDS prevalence estimators (Mills et al., 2014; Rocha, Thorson, Lambiotte, & Liljeros, 2016; Sperandei et al., 2018). However, almost all currently proposed estimators for RDS samples aim only to estimate the prevalence of a condition in the population of interest and not the identification of factors associated with that condition. In order to address this, Bastos et al. (2018) proposed a model-based estimator, called RDS-B, which can be used to estimate both prevalence and associated factors. Notwithstanding the capacity of RDS-B to estimate associated factors, the authors only used the estimator in its simplest form to estimate prevalence and did not fully address its model-based characteristics.

Several researchers who have analyzed RDS-based datasets have applied simple logistic regression estimators to assess the putative association between covariates and outcomes, irrespective of the varied study designs and the very characteristics of RDS, especially the underlying network structures (e.g., Do et al., 2018; Liu et al., 2018; Toro-Tobón, Berbesi-Fernandez, Mateu-Gelabert, Segura-Cardona, & Montoya-Vélez, 2018). Conversely, others try to use some form of weighted logistic regression, adding weights obtained from reported network sizes (e.g., Hotton, Quinn, Schneider, & Voisin, 2018; Ndori-Mharadze et al., 2018; Szwarcwald et al., 2018). However, the influence of such sampling weights has not been assessed beyond what has been defined as the basic diagnostic tools to double-check either the sound or improper use of the standard RDS procedures (e.g., Gile et al., 2015).

The purpose of this paper is to address this gap in knowledge by assessing the performance of three logistic regression models in estimating true, expected relationships when applied to RDS samples generated by simulations. These estimators were then applied to a real-life RDS sample data of transgender women from a large Brazilian study with a sample of 2,846 participants.

**METHODS**

**Simulation**

A total of four connected populations (N=10,000) were simulated using two random graph models, with and without the simulation of nested subpopulations. The random graphs used and the main parameters for each population were as follows:

- *Erdös-Rényi without subpopulations (ER1)*: the simplest random graph structure, initially proposed by Erdös and Rényi (1959), where links between two members of the population were established at random, with a fixed probability (P). P was set at 0.0025;

- *Erdös-Rényi with nested subpopulations (ER2)*: this population is similar to the previous (ER1). However, instead of one population, five subpopulations were nested within the P set at 0.0125. Only ten individuals in each of the five subpopulations were allowed to connect with other subpopulations. They were chosen at random.

- *Barabasi-Albert without subpopulations (BA1)*: the scale-free model created by Barabasi and Albert (1999), also known as the "richer get richer", follows a power-law distribution for connectivity. In summary, the population starts with one individual and every new individual entering the population has the probability of linking with old members proportionally to the connectivity degree (i.e. number of contacts) of each individual. It generates few individuals with extremely high connectivity degrees and the majority of the population with few connections. The parameter needed is the number of links each new individual will establish when joining the population. In this simulation, such links were set to 12.5.

• *Barabasi-Albert with nested subpopulations (BA2)*: five subpopulations with 2,000 individuals each were generated to construct this population. Subsequently, ten individuals from each subpopulation with the highest degree were chosen to link randomly across these subpopulations.

All parameters were set in order to obtain, whatever the model, a mean connectivity degree of 20 in all populations.

**Explanatory Variables**

To assess the performance of logistic estimators emulating actual associations, two binary explanatory variables were added as attributes of each individual, apart from the infected/not-infected status (see infection process below). They were named E1 and E2. Each one presents 50% of positive and negative cases, randomly distributed in the population. During the infection process, each individual in the population with a positive E1 variable will present twice the chance of being infected. The purpose of this is to force a statistical association between E1 and the disease, while the variable E2 will present no relationship with disease.

**Infection Processes**

Four infection processes were simulated to emulate the dissemination of a particular disease in each of the populations. All processes are variations of the classical Susceptible-Infected (SI) model, where the infected individual does not recover from the disease. In the first process, individuals were selected at random and defined as "infected". The other three processes were dependent on network contacts; all three started with some randomly selected individuals defined as "infected" but, unlike the first process, from there the infection followed through the network contacts in successive waves. In each wave, all individuals connected to the infected ones had a probability of 0.005 to be infected. This infection rate was selected to avoid an out of control increase of the infected population (i.e. an unexpected outbreak). Each newly infected individual could infect their contacts in subsequent waves. All infected individuals kept infecting their contacts until the desired prevalence was reached. The infection prevalence was set at 30%.

Processes started with 10, 100, and 500 infected individuals, creating infections dependent on network connectivity. In the case of 10 initially infected individuals, all those infected were more closely related to the network of the initial individuals, given each individual would generate, on average, an infected tree of about 300 individuals. In the case of 500 initially infected individuals, there would be a lower network connectivity dependency, with expected trees of only six individuals each. Also, the random process can be considered a particular case, where the process starts with 3,000 infected individuals (prevalence = 30% of 10,000). These processes simulate diseases that depend on interaction between susceptible and infected individuals.

**Sampling Process**

Benchmark samples were obtained in a simple random process, applied to each combination of population versus simulated infection pattern.

RDS samples were obtained simulating an RDS process. All RDS sampling processes were launched using three randomly selected individuals ("seeds"). Each seed recruited randomly from their network one to three contacts, with probabilities of 0.40, 0.40, and 0.20, respectively. These probabilities were based on empirical data from a study with drug users from Belo Horizonte, Brazil (unpublished data). Each recruited individual repeats the process, recruiting additional individuals from their network, and this pattern is repeated until the desired sample size was obtained. It is essential to highlight that, although similar to the infection process previously described, each individual in the population recruits only one to three individuals. In contrast, in the infection process, they keep infecting other individuals until the end of the process.

No homophily-related bias was explicitly incorporated into the recruitment process, although previous studies have suggested that homophily may influence the process (Gile et al., 2015). The simulated samples were designed to reproduce a "perfect world", following the RDS method assumptions, that is, seeds are recruited randomly, each recruiter recruits randomly among their contacts, no recruitee refuses to participate and all report their network size accurately.

In all cases, 1,000 samples with three sample sizes (i.e. 100, 250, and 500 individuals) were obtained from each combination of population and infection, and applied to all three logistic estimators.

**Logistic Estimators**

Three variations of logistic regression estimators were applied to the above-simulated data. For each, a model with both variables and interaction was fitted.

The first, used on both RDS and random samples, was the logistic regression estimator (Sperandei, 2014), with the frequentist likelihood estimator. It will be named here the "unweighted logistic", given the other two estimators are weighted.

The second type of regression, called here "RDS-weighted logistic", takes into consideration the study design and weightings of each individual using the same form of weighting used in RDS-I and RDS-II estimators (Heckathorn, 1997, 2002; Salganik & Heckathorn, 2004). It weighs results from the simulations proportionally to the inverse of the reported degree of each individual (Volz & Heckathorn, 2008).

The third type of regression estimator, called "RDS-B" (Bastos et al., 2018), is a Bayesian version of the RDS-weighted logistic, where weakly informative priors are set to the coefficients (Gelman, Jakulin, Pittau, & Su, 2008), and the weighted likelihood, called pseudo-likelihood, is combined with the prior using Bayes theorem, leading to the pseudo-posterior distribution (Savitsky & Toth, 2015). Posterior means were used in order to make a comparison among estimators, and 95% credible intervals were used to represent uncertainty.

In the case of randomly selected samples, only the unweighted logistic estimator was used, defining a benchmark performance.

**Performance Assessment**

The performance assessment was accomplished by the observed coverage metric, also known as coverage probability (Dodge, 2003). This is the proportion of times the confidence interval of each estimator contains the populational parameters simulated. It means that, for the coefficient of E1, the OR confidence interval contains the parameter 2 simulated for each population. For this coefficient, the confidence interval also needs to exclude the value of 1, meaning a significant coefficient. The rationale for this second criterion is to avoid too wide confidence intervals being considered a good performance. For the coefficients of E2 and the interaction E1xE2, the OR confidence interval must contain the value of 1, meaning a non-significant interval, which is the simulated situation. For these two coefficients, the complementary probability (1 - coverage) will be used as an estimate of type-I error probability. Finally, a combination of E1, E2, and interaction results will be built to investigate the probability of a combined correct estimation from the model, meaning a significant E1 coefficient and non-significant coefficients for E2 and interaction E1 x E2. The word "significant" here was used in a broad sense, related to the usual 95% confidence interval, although we acknowledge that in Bayesian models these definitions are not strictly adequate.

All performances were compared to the random samples' performance for each combination of population and infection.

**Real-Life Data**

All four estimators were subsequently applied to the Divas Research dataset (Bastos et al., 2018), which is a large RDS-based study conducted across 12 cities in Brazil that collected data on 2,846 transgender women.

The entire dataset was combined and considered as one population, from where the expected parameters were estimated. Four variables were considered in this study to assess the performance of the estimators. HIV status (positive x negative) was considered the main outcome. The two explanatory variables considered were whether the person had acted as a sexual worker anytime in their life (explanatory variable 1 – E1) and whether the person had moved from their place of birth anytime during life (E2). E1 is expected to be related to HIV status, while E2 is not. The fourth variable was the reported number of contacts (network degree), which was used in RDS estimations. A total of 2,548 individuals were used to avoid missing information in any of the variables considered.

From this population, samples were extracted with sizes of 100, 250, and 500 individuals. First, 1,000 random samples of each sample size were used as benchmarks, similar to what was done in the simulation. Second, 1,000 samples were drawn following the RDS process. As the objective here is to observe the impact of real-world constraints and bottlenecks in the sampling procedure, these samples were extracted respecting the original RDS sampling from the dataset. Real seeds were randomly selected and the original recruitment trees were followed from each seed until the desired sample size was reached. By doing this, each sample used was a subsample of the original dataset, presenting all the characteristics found in real-life sampling.

Again, similarly to the process used in the simulation, the sample results were compared to the observed result from the population, and the number of correct estimations was counted.

The Divas study received ethics approval from the Escola Nacional de Saúde Publica (CAAE 49359415.9.0000.5240). All participants signed an informed consent form to take part in the study. The dataset was provided in an unidentified form and no additional approval was necessary for the current study.

All simulations and analyses used R software, version 3.4.4 (R Core Team, 2018) and its packages *igraph* (Csardi & Nepusz, 2006), *survey* (Lumley, 2004), and *arm* (Gelman & Su, 2018).

**RESULTS**

Results of the simulated populations can be seen in Figure 1. Red dots represent infected individuals, while blue dots are non-infected individuals. A considerably different pattern can be noted between the two random graph models used and an even more dramatic effect between clustered and non-clustered populations. Comparing ER and BA networks, it is clear that highly connected individuals, located on the borders of the population, have a higher chance of becoming infected in the BA model. In ER models, as the distribution of degrees does not present heavy tails, the infection is more uniformly spread. The same pattern can be observed in models with subpopulations well defined, with one additional characteristic: the clustered nature of these models resulted in parts of the population being almost untouched by infection.

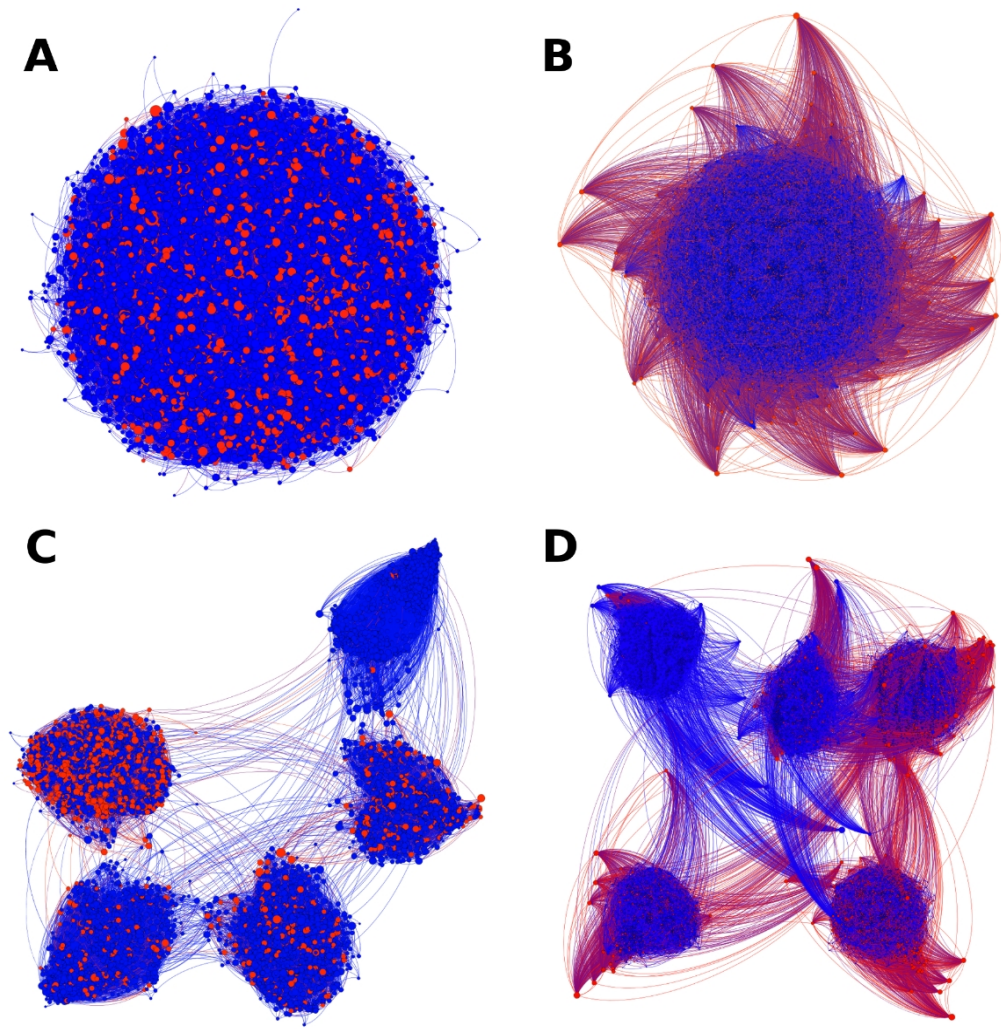

**Figure 1.** Populations created. Blue vertices and edges are for non-infected individuals. Red vertices and edges are for infected individuals. A: ER1 model. B: BA1 model. C: ER2 model. D: BA2 model.

Table 1 presents the main characteristics of each simulated population as well as the Divas dataset. It can be noted that all main characteristics were successfully simulated. The Barabasi-Albert models showed a discrepancy between the average and the median degree due to the asymmetric nature of the model degree's distribution.

**Table 1.** Main characteristics of simulated and Divas populations.

| Characteristic | Population | | | | |
|---|---|---|---|---|---|
| | ER1 | ER2 | BA1 | BA2 | Divas |
| Mean Degree | 20.03 | 19.95 | 19.99 | 19.95 | 20.21 |
| Median Degree | 20.0 | 20.0 | 14.0 | 14.0 | 10.0 |
| Min – Max Degree | 4 – 37 | 5 – 39 | 10 – 541 | 10 – 247 | 2 – 100 |
| Infection Prevalence (%) | 30.2 – 31.6* | 30.4 – 31.9* | 30.0 – 33.0* | 29.8 – 32.5* | 29.98 |
| E1 Prevalence (%) | 50 | 50 | 50 | 50 | 76.4 |
| E2 Prevalence (%) | 50 | 50 | 50 | 50 | 60.9 |
| E1 Odds Ratio | 1.97 – 2.00* | 1.98 – 2.05* | 1.97 – 2.05* | 1.98 – 2.04* | 1.83 |
| E2 Odds Ratio | 0.90 – 1.01* | 0.85 – 1.10* | 0.76 – 1.09* | 0.83 – 1.04* | 1.26 |
| E3 Odds Ratio | 0.95 – 1.33* | 0.94 – 1.21* | 0.92 – 1.45* | 0.98 – 1.40* | 1.31 |

\* Values represent the minimum and maximum range across the four types of infection

The simulated prevalence ranged from 14.6% (ER2) to 17.2% (BA1), very close to the desired value (15%). Regarding true ORs observed in the population, general logistic models fitted to the whole population (one for each population) detected significant ORs for variable E1, all between 1.95 and 2.05, after adjusting for E2 and the interaction. For variable E2, true ORs ranged from 0.81 to 1.10, all of them non-significant, as expected. Lastly, for the interaction factor (E1xE2), true ORs varied from 0.90 (ER1) to 1.20 (BA2). These results confirm the simulation process was adequate. Regarding the Divas population, a pattern towards a power-law distribution of connectivity and a clustered behaviour is expected, given the way the population was created, joining samples from twelve cities. This means that no individual will recruit out of their own city. Overall, the Divas dataset was most similar to the BA2 simulated population.

Figure 2 presents observed coverage probability results according to the network model, infection process, sample size, and estimators used for coefficient E1 alone. The most evident effect was related to the sample size. The higher the sample size, the higher the coverage. Regarding estimators themselves, three of them had similar performances, with slightly better performance by the traditional logistic estimator applied to RDS samples. The estimator with the worst performance was the weighted-logistic estimator. However, even this estimator did not perform substantially below the logistic estimator applied to random samples (benchmark) and could be considered a satisfactory estimator. In regards to the effect of network models, it can be observed that populations without heavy tails in the distribution of degrees (ER1 and ER2) present very small difference between estimators, while heavy tail distributions of degree inside the population (BA1 and BA2) seems to affect heavily the weighted estimators (RDS and Bayes) and favor the unweighted estimator applied to RDS samples. The presence of subpopulations (ER2 and BA2) had little to no effect on the estimators' performance for E1 or the analysis of the combined coefficients. Lastly, it is interesting to note that, in Barabasi-Albert model-based populations, the unweighted estimator applied to RDS samples presented a better performance when the infection was not random even when compared to random samples.

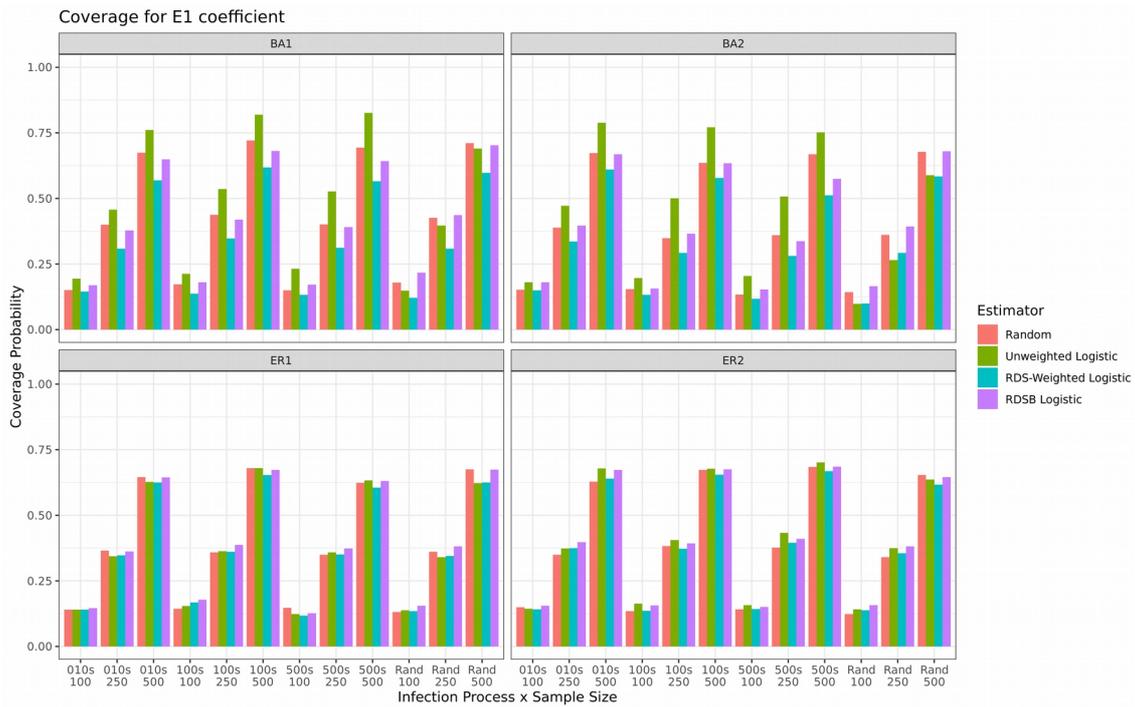

**Figure 2.** Observed coverage probability results according to the combination of network models (each subgraph, as labelled), sample size (100, 250, 500) and infection process (10s, 100s, 500s, Rand).

In relation to type-I error probability, Figures 3 and 4 present the results for E2 and interaction coefficients, respectively. Irrespective of the type of infection, sample size, network model or estimator, the type-I error probability for both coefficients was close to the expected value of 5%. Only for BA networks, under random infection, with n=500 (and to a lesser extent with n=250), the error rate was above this threshold, especially for the unweighted estimator applied to RDS samples. The error rate for the combined coefficients shows a general trend for an addictive effect, showing a certain independence between the coefficients error (Figure 5).

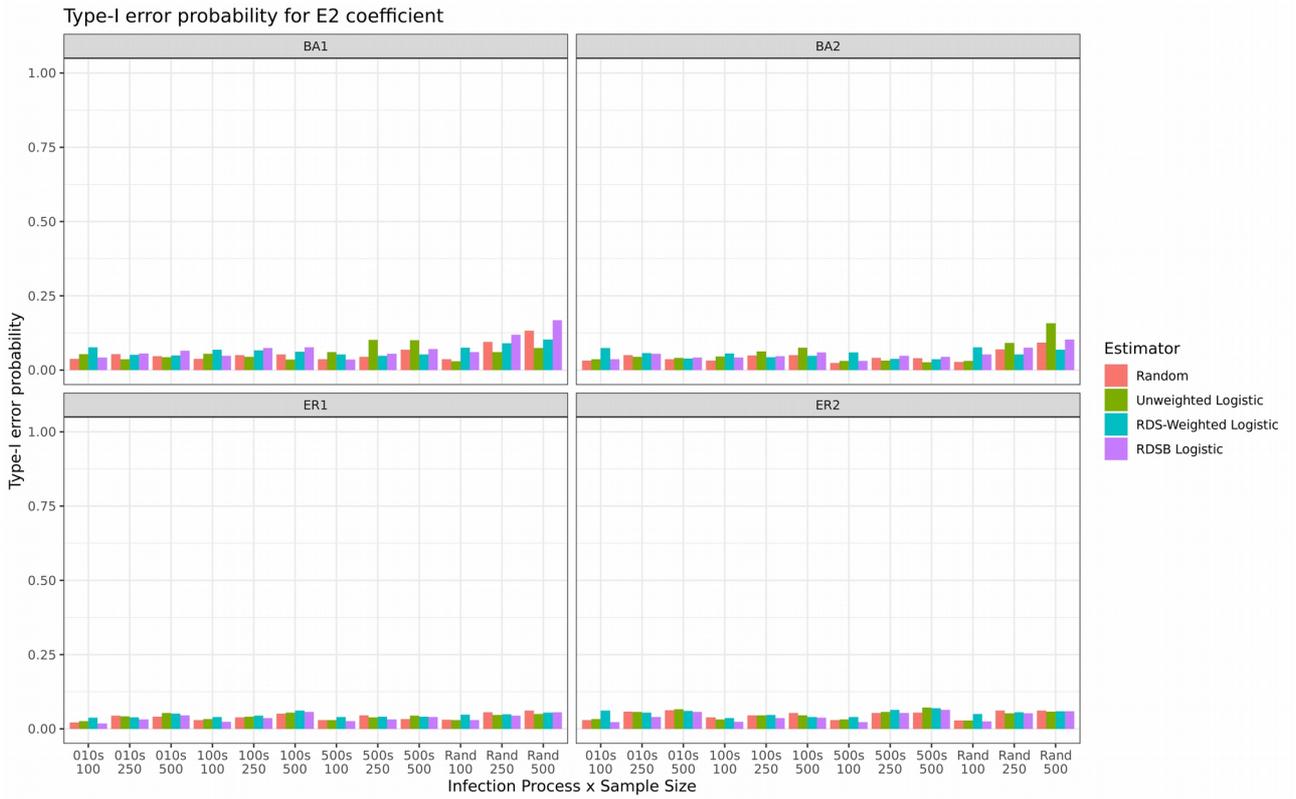

**Figure 3.** Type-I error rate for the E2 coefficient according to network models (each subgraph, as labelled), sample size (100, 250, 500) and infection process (10s, 100s, 500s, Rand).

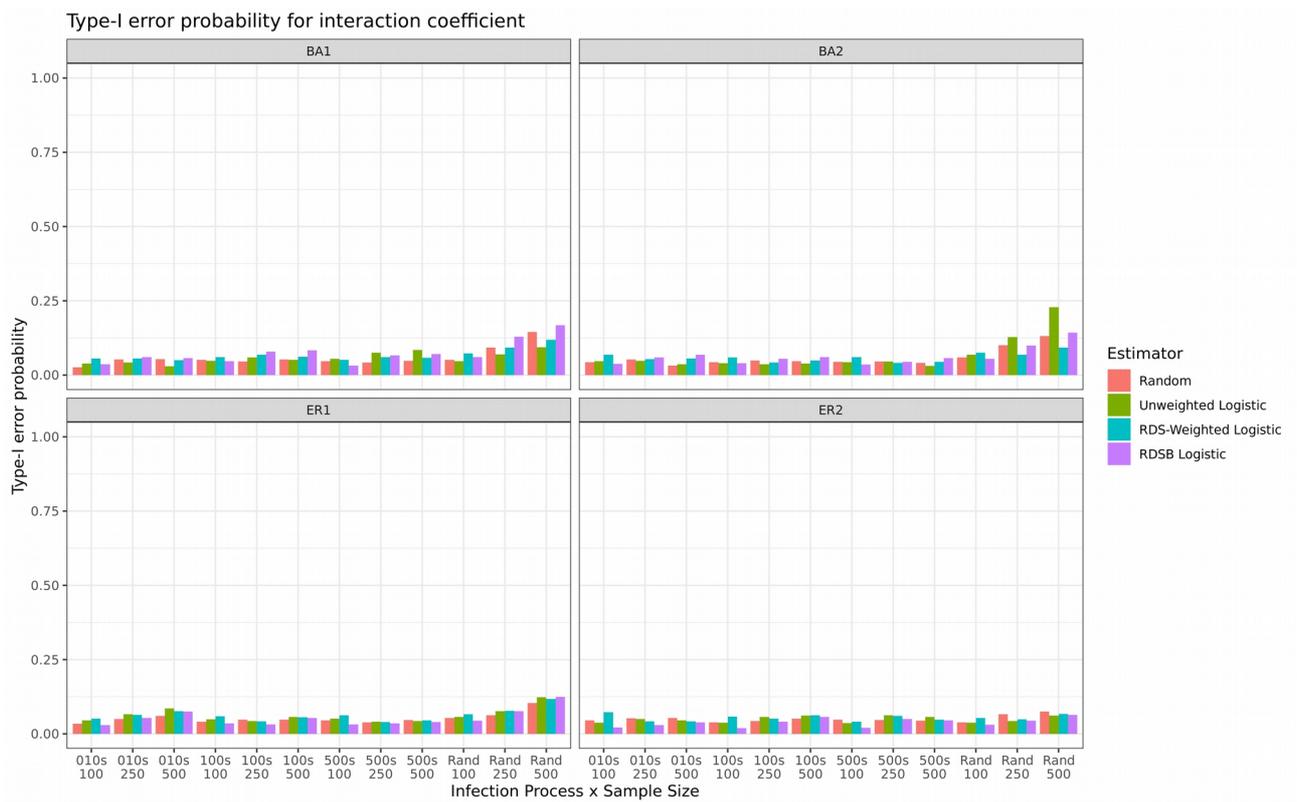

**Figure 4.** Type-I error rate for the interaction coefficient according to network models (each subgraph, as labelled), sample size (100, 250, 500) and infection process (10s, 100s, 500s, Rand).

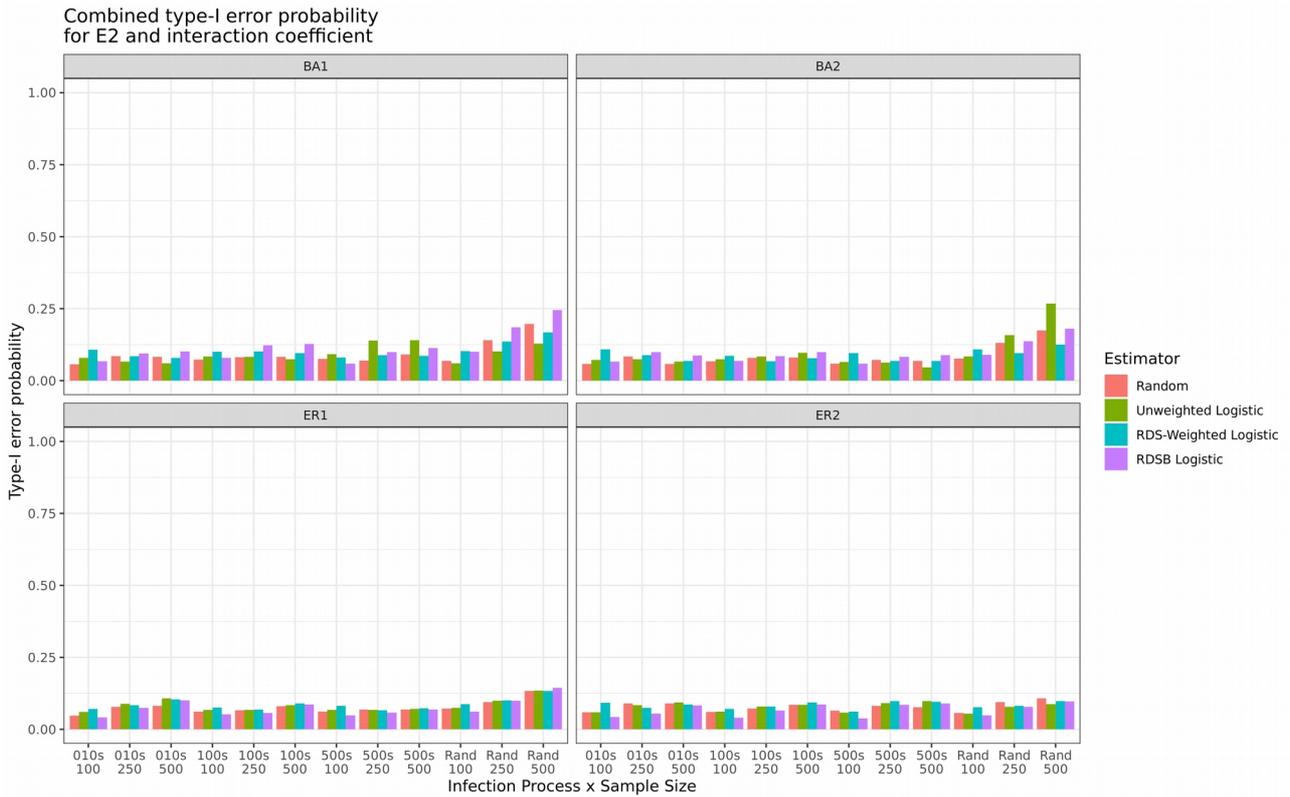

**Figure 5**. Type-I error rate for the E2 and the interaction coefficients according to network models (each subgraph, as labelled), sample size (100, 250, 500) and infection process (10s, 100s, 500s, Rand).

When the analyses of all three coefficients are combined, it is possible to notice the general performance of the estimators to find the "right answer" from the samples: a significant E1 coefficient with a confidence interval containing the simulated E1 effect plus non-significant E2 and interaction coefficients. Figure 6 illustrates how results are very similar to those for the E1 coefficient, given the general stability of E2 and interaction results. The results for the random infection were the most affected, especially by the higher type-I error rate.

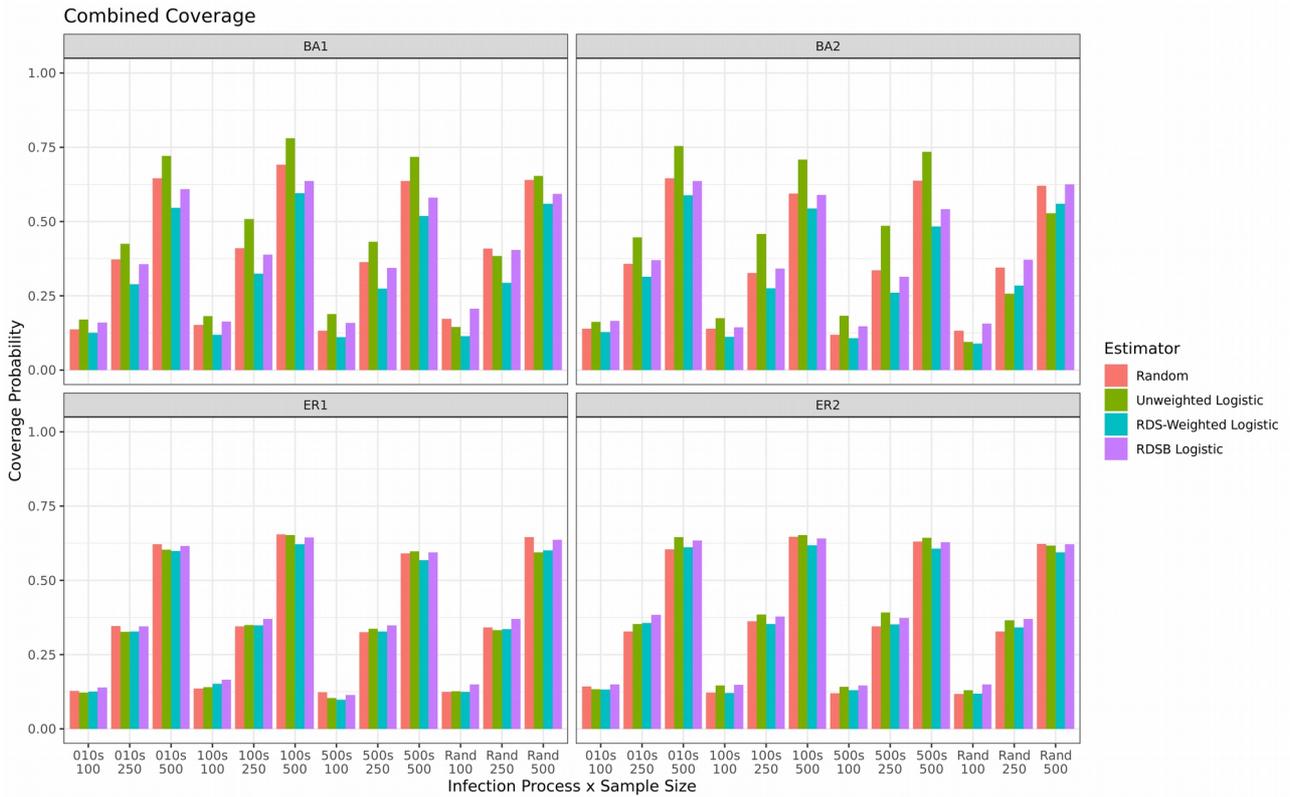

**Figure 6.** Observed coverage probability results for the combination of coefficients according to network models (each subgraph, as labelled), sample size (100, 250, 500) and infection process (10s, 100s, 500s, Rand).

A more interesting result was observed when the estimators were applied to the Divas dataset. First, the random samples behaved as expected, with a proportional increase in coverage for the E1 coefficient according to the sample size (Figure 7). Second, the unweighted estimator presented a similar behavior when applied to RDS samples compared to random samples. Third, weighted estimators presented a somewhat strange behavior, with unusual high coverage for smaller samples (compared to random), and smaller improvements with increasing size, especially the RDS-B, which demonstrated a drop when the sample reached 500 individuals. This pattern was the same for the combination of all coefficients.

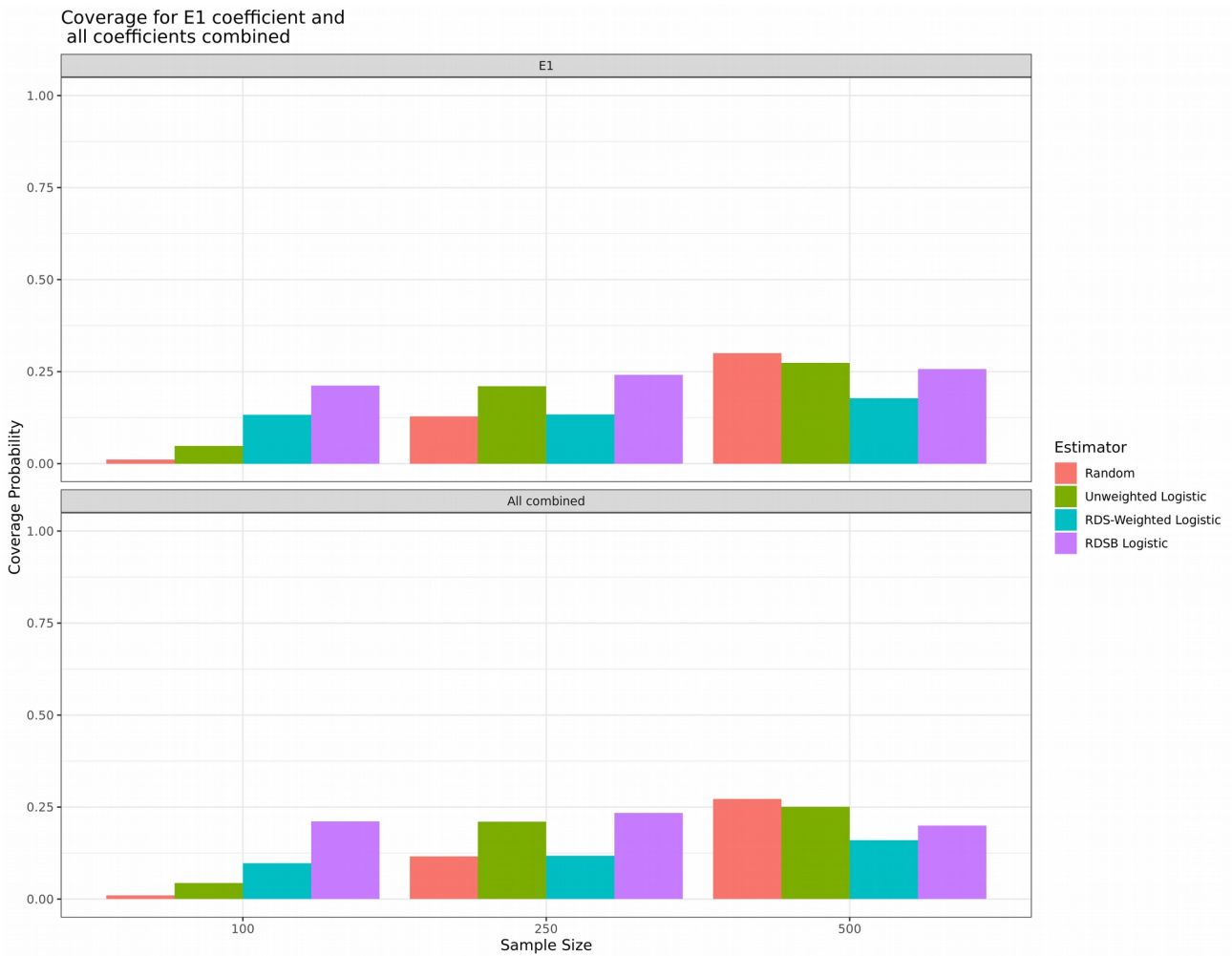

**Figure 7.** Observed coverage probability for E1 and all coefficients combined according to sample size and estimator.

When looking at the type-I error rate (Figure 8), they were well below the expected for the sample size of 100 and around 5% for the unweighted logistic estimator, either applied to random or RDS samples. The weighted estimators showed a higher error rate, especially for the RDS-weighted logistic estimator, which reached more than 40% with sample size of 100. This represents a very high probability of wrong results when using this estimator.

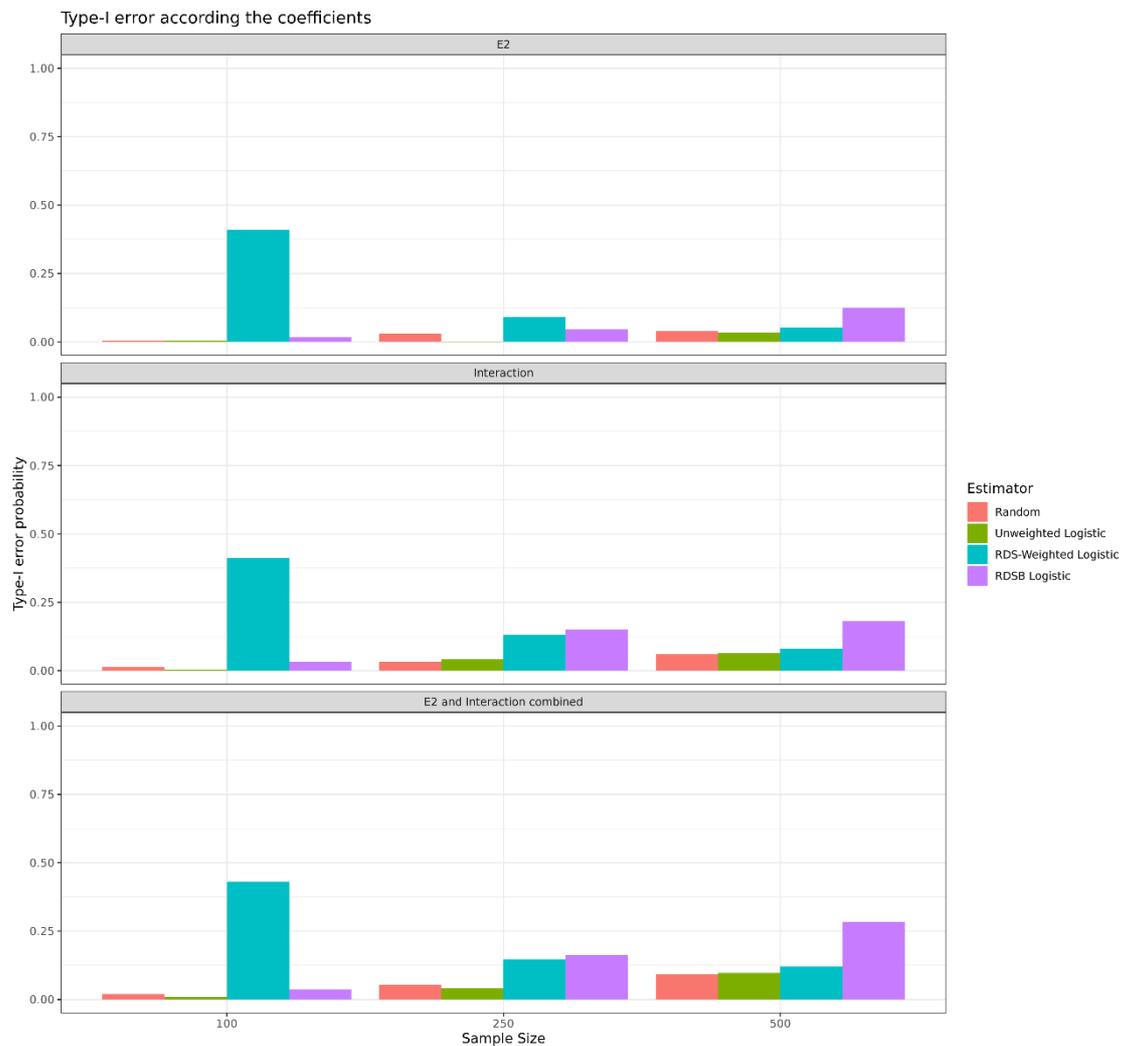

**Figure 8.** Type-I error rate for the E2 and the interaction coefficients according to sample size and estimator.

DISCUSSION

The RDS method has been widely used and recommended as a sampling method to recruit hard-to-reach populations, such as drug users, sex workers, transgender individuals, among others. Although its ability to find and recruit members of these "hidden" populations is uncontroversial, its use as an estimator method is still disputed (Sperandei et al., 2018). Moreover, the use of model-based estimators to study relationships between response and explanatory variables has been poorly assessed, especially in regards to the basic question of when to use sampling weightings (Schonlau & Liebau, 2012). These issues notwithstanding, researchers have used traditional logistic estimators or some form of weighted logistic applied to RDS samples. A quick survey of the Pubmed database identified 70 studies published between 2018 and 2019 applying logistic regression models to RDS samples, with 48.6% (n=34) using unweighted estimators, 44.3% (n=31) using some form of weighting with network degrees, and 7.1% (n=5) presenting both weighted and unweighted models. This pattern highlights the evident lack of consensus in the current literature on which type of estimator should be used.

Our simulations have demonstrated not only the impact of data and population characteristics but also the estimator used on results of an RDS study. Although some interactions with other factors must be considered, it seems that weighted and unweighted estimators performed relatively well when compared to logistic regression applied to random samples.

To the best of our knowledge, only one study assessed the impact of weights used on RDS sample estimates from logistic regression models (or any other form of model estimates) using a simulation approach, and, similarly to our results, it concluded that unweighted estimators perform better than the weighted ones (Avery et al., 2019). However, the lack of a clear structure in the simulated network connections and the absence of real data to reflect real sampling problems, in comparison to "perfect" simulated samples, left many issues unaddressed. First, Avery et al.'s (2019) study used only simple logistic models, with just one explanatory variable, not considering the effect of interaction between explanatory variables on the result. Second, this study confounded clustering with homophily, when they are, in fact, separate concepts (Rocha et al., 2016; Sperandei et al., 2018). Clustering represents the phenomenon of individuals being more connected to their similar ones (in one or more characteristics such as age, geography, etc.), whereas homophily relates to preferential recruitment, where people choose to recruit those peers with particular characteristics (that the recruiter also possesses), instead of recruiting randomly (Lu et al., 2012). In the present study, we addressed these limitations by creating populations based on theoretical graph models, controlling the connectivity process. From the results, comparing the two models used here, it is clear the impact of the nature of connectivity on the performance of estimators, which is reinforced by previous research on simple prevalence estimators (Rocha et al., 2016; Sperandei et al., 2018).

In addition, we used an adapted concept of "coverage probability" to reflect not only the identification of correct estimation of the E1 coefficient but also the simultaneous identification of E2 and the E1xE2 interaction, representing the proportion of correct estimation for the complete hypothesis. It represents a more restrictive criterion compared to the usual coverage because it requires all three hypotheses (E1, E2, E1xE2) being true at the same time.

However, simulations can only approximate the characteristics of the real world, their success being dependent on previous knowledge about the population being simulated. This knowledge, in the case of hard-to-reach populations, can be very restricted. The use of real data allows us to observe what happens when RDS is applied in the real world. In our simulated scenarios, RDS sampling followed best practice described for the method, with random selection of seeds, long recruitment trees, and each recruiter "selecting" randomly amongst their peers (Salganik & Heckathorn, 2004; Volz & Heckathorn, 2008). In practice, it is common to see "dead seeds" (seeds that do not recruit any peers), recruitment trees with mixed length, and true homophily, with recruiters choosing selectively amongst their peers (Li et al., 2018). Also, time, resource, and logistical constraints are common, and their impacts on estimation are unknown (Truong et al., 2013, Valois-Santos et al., 2020). Considering a large sample as a population, and using real recruitment trees as RDS samples, is not a perfect approach. However, we argue that it is one of the best possible ways of assessing RDS estimators in real life.

In this dataset, the random samples acted as a benchmark to what would be expected, given that, for any population, random samples are considered the gold standard sampling method. The results show the expected increase in coverage according to the sample size. The most exciting finding was the performance of the unweighted logistic estimator applied to RDS samples, which showed similar results compared to random samples, sometimes even better. The results with real data represent a decreased performance in comparison to simulation results, showing the effects of differences between theoretical sampling procedures and real ones; however, it still performs well and is a good alternative to be used with RDS samples, similarly to what Avery et al. (2019) found.

On the other hand, weighted estimators presented more aberrant behavior, especially the RDS-B, which presented higher coverage with smaller samples. At a lower intensity, the RDS-weighted estimator also showed an unexpectedly high power with the 100 samples, but the increase with bigger sample sizes was not so considerable. This behavior, also partially observed in the performance of unweighted logistic, is probably related to the differences in simulated and real sampling procedures. In relation to the type-I error rate, the RDS-weighted estimator showed a very high result, representing a big chance of a wrong result.

Several studies have demonstrated the advantages of weighting procedures for the simple prevalence of RDS estimators (Goel & Salganik, 2010; Mills et al., 2014; Sperandei et al., 2018). However, the present results demonstrate that weighting may not be the best option when it comes to regression coefficient estimates, making the unweighted estimator the preferable one instead.

## CONCLUSION

In summary, this study demonstrated how unweighted logistic regression is the best option to be used with RDS samples, particularly when the requirements of the RDS method are respected. However, even in real RDS samples, it achieved a performance as good as the random sampling performance. These findings suggest that the RDS method is applicable to a broader spectrum of research designs, even where true random sampling is difficult to be achieved.